\documentclass[manuscript]{aastex}
\usepackage{lineno}
\linenumbers

\usepackage{graphicx}
\usepackage{natbib}
\usepackage{upgreek}
\usepackage{color}

\newcommand{\kep}{$K_{ep}$}
\newcommand{\nh}{\bar{n}_{\rm H}}

\newcommand{\hi}{H$\,\scriptstyle{\rm I}$~}
\newcommand{\halpha}{H$\alpha$~}

\slugcomment{To be submitted to ApJ: v6.8,  \today}

\shorttitle{\emph{Fermi} LAT Observations of the Cygnus Loop}
\shortauthors{\emph{Fermi} LAT collaboration}

\begin{document}

\title{\emph{Fermi} Large Area Telescope Observations of the Cygnus Loop Supernova Remnant}

\author{
H.~Katagiri\altaffilmark{1,2}, 
L.~Tibaldo\altaffilmark{3,4,5,6,7}, 
J.~Ballet\altaffilmark{5}, 
F.~Giordano\altaffilmark{8,9}, 
I.~A.~Grenier\altaffilmark{5}, 
T.~A.~Porter\altaffilmark{10}, 
M.~Roth\altaffilmark{11}, 
O.~Tibolla\altaffilmark{12}, 
Y.~Uchiyama\altaffilmark{10}, 
R.~Yamazaki\altaffilmark{13}
}
\altaffiltext{1}{College of Science, Ibaraki University, 2-1-1 Bunkyo, Mito, Ibaraki 310-8512, Japan}
\altaffiltext{2}{email: katagiri@mx.ibaraki.ac.jp}
\altaffiltext{3}{Istituto Nazionale di Fisica Nucleare, Sezione di Padova, I-35131 Padova, Italy}
\altaffiltext{4}{Dipartimento di Fisica ``G. Galilei'', Universit\`a di Padova, I-35131 Padova, Italy}
\altaffiltext{5}{Laboratoire AIM, CEA-IRFU/CNRS/Universit\'e Paris Diderot, Service d'Astrophysique, CEA Saclay, 91191 Gif sur Yvette, France}
\altaffiltext{6}{Partially supported by the International Doctorate on Astroparticle Physics (IDAPP) program}
\altaffiltext{7}{email: luigi.tibaldo@pd.infn.it}
\altaffiltext{8}{Dipartimento di Fisica ``M. Merlin'' dell'Universit\`a e del Politecnico di Bari, I-70126 Bari, Italy}
\altaffiltext{9}{Istituto Nazionale di Fisica Nucleare, Sezione di Bari, 70126 Bari, Italy}
\altaffiltext{10}{W. W. Hansen Experimental Physics Laboratory, Kavli Institute for Particle Astrophysics and Cosmology, Department of Physics and SLAC National Accelerator Laboratory, Stanford University, Stanford, CA 94305, USA}
\altaffiltext{11}{Department of Physics, University of Washington, Seattle, WA 98195-1560, USA}
\altaffiltext{12}{Institut f\"ur Theoretische Physik and Astrophysik, Universit\"at W\"urzburg, D-97074 W\"urzburg, Germany}
\altaffiltext{13}{Department of Physics and Mathematics, Aoyama Gakuin University, Sagamihara, Kanagawa, 252-5258, Japan}


\begin{abstract}
We present an analysis of the gamma-ray measurements by the Large Area Telescope~(LAT) onboard
the \textit{Fermi Gamma-ray Space Telescope} in the region of
the supernova remnant~(SNR) Cygnus Loop~(G74.0$-$8.5). 
%
We detect significant gamma-ray emission 
associated with the SNR in the energy band 0.2--100~GeV.
The gamma-ray spectrum
shows a break in the range 2--3~GeV. The gamma-ray luminosity is $\sim$~$1 \times
10^{33}$erg~s$^{-1}$ between 1--100~GeV, much lower than those of other GeV-emitting SNRs.
The morphology is best represented by a ring shape, with inner/outer radii
 0$^\circ$.7~$\pm$~0$^\circ$.1 and 1$^\circ$.6~$\pm$~0$^\circ$.1.
Given the association  among X-ray rims, \halpha filaments and gamma-ray emission,
we argue that gamma rays originate in interactions between particles accelerated in the
SNR and interstellar gas or radiation fields adjacent to the shock regions.
The decay of neutral pions produced in nucleon-nucleon interactions between accelerated hadrons and
interstellar gas provides a reasonable explanation 
for the gamma-ray spectrum. 

\end{abstract}


\section{Introduction}
Diffusive acceleration by supernova shock waves can  
accelerate particles to very high energies~\citep[e.g.,][]{blandford87}.
Gamma-ray observations are a useful probe of these mechanisms complementary to other wavebands.
So far, observations
 by the Large Area Telescope~(LAT) on board 
the \textit{Fermi Gamma-ray Space Telescope}
have demonstrated that bright gamma-ray sources coincident with middle-aged supernova remnants~(SNRs)
interacting with dense molecular clouds~\citep{LAT-W51C, LAT-W44, LAT-IC443, LAT-W28, LAT-W49B} exhibit steep gamma-ray spectra above a few GeV.
A possible conventional explanation
for these spectral properties is
that the energy distribution of cosmic rays (CR) is greatly influenced by their diffusive transport~\citep[e.g.,][]{CRtransport1,CRtransport2,CRtransport3,CRtransport4}.
On the other hand, these features can also be explained by reacceleration of pre-existing cosmic rays at a cloud shock and subsequent adiabatic compression where strong ion-neutral collisions accompanying Alf$\acute{\rm v}$en wave evanescence lead to a steepening of the spectrum of accelerated particles~\citep{UchiyamaModel}.
Furthermore, using three-dimensional magnetohydrodynamic simulations, \cite{twostep_acceleration}
show that the interaction between a supernova blast wave and inhomogeneous interstellar clouds
formed by thermal instability generates multiple reflected shocks, which can further energize
cosmic-ray particles originally accelerated at the blast-wave shock and produce the spectral break.
Since the gamma-ray bright regions are expected to be different in the aforementioned models,
studying SNRs with large apparent sizes can help to disentangle the origin of the spectral features.

The Cygnus Loop~(G74.0$-$8.5) is one of the most famous and well-studied middle-aged SNRs. 
The size~($\sim$~3$^\circ$) makes it an ideal candidate
for detailed morphological studies in high-energy gamma rays since it is larger than the LAT angular resolution above a few hundred MeV.
The large angular offset from the Galactic plane~($b \sim -8^\circ.5$)
 reduces the problems of background contamination and permits a detailed
study of the environment around the shock region by means of infrared/optical/UV observations.
The shell-like X-ray emission from thermal plasma is prominent in the northern region of the
remnant, with a
blowout in the southern rim~\citep{EinsteinXray}. The radio spectrum from the
limb-brightened shells is non-thermal~\citep{HistoricalRadio}.
No correlation with dense molecular clouds has been reported, although blast waves on the
western limb might encounter molecular material~\citep{COwestern}.
The distance from the Earth is estimated to be $540^{+100}_{-80}$~pc based on the proper motion of optical
filaments
in conjunction with models of non-radiative shocks~\citep{Distance}.
The age has been estimated to be $\sim$~$2 \times 10^4$~yr based on plasma parameters derived from X-ray data~\citep{age2} and $\sim$~$1.4 \times 10^4$~yr based on the shock model and X-ray measurements~\citep{age3}.

Four LAT sources positionally associated with the Cygnus Loop SNR
are listed in the 1FGL catalog~\citep{1yrCatalog}.
In this paper, we report a detailed analysis of \emph{Fermi} LAT observations in the Cygnus Loop region.
First, we briefly describe the observations and data selection in
Section~\ref{sec:obs}.
The analysis procedure and the results are presented in Section~\ref{sec:ana}, 
with the study of the morphology and spectrum of emission associated with the Cygnus Loop. 
Results are then discussed in Section~\ref{sec:discuss} 
and our conclusions are presented in Section~\ref{sec:conclusion}.

\section{OBSERVATIONS AND DATA SELECTION}
\label{sec:obs}
The LAT is the main instrument on \emph{Fermi}, detecting gamma rays from
$\sim$~20~MeV to $>$~300~GeV\footnote{ 
As noted below in the present analysis we use only events with energies $>$~200~MeV.}.
Details about the LAT instrument and pre-launch expectations for the performance can be found in
\cite{Atwood09}. 
Compared to earlier high-energy gamma-ray telescopes, the LAT has a larger field of
view~($\sim$~2.4~sr), a larger effective area~($\sim$~8000~cm$^2$ for $>$1~GeV on-axis) and
improved point-spread function~(PSF; the 68\% containment angle $>1$~GeV is smaller than 1$^\circ$). 

Routine science operations with the LAT began on August 4, 2008.
We have analyzed events in the region of the Cygnus Loop collected from August 4, 2008, to August 1,
2010,
with a total exposure of $\sim$~6~$\times$~10$^{10}$~cm$^2$~s~(at 1~GeV).
The LAT was operated in sky-survey mode for almost the entire period.
In this observing mode the LAT scans the whole sky, obtaining complete sky coverage
every 2
orbits~($\sim$~3~hr) and approximately uniform exposure.

We used the standard LAT analysis software, the \emph{ScienceTools} version v9r16,
publicly available from the \emph{Fermi} Science Support Center (FSSC)\footnote{Software and
documentation of the \emph{Fermi} \emph{ScienceTools} are distributed by \emph{Fermi} Science
Support Center at http://fermi.gsfc.nasa.gov/ssc}, and applied the following event selection
criteria: 
1) events have the highest probability of being gamma rays, i.e., they are classified in the
so-called Pass 6 \emph{Diffuse} class \citep{Atwood09}, 
2) events have a reconstructed zenith angle less than 105$^\circ$, 
to minimize the contamination from Earth-limb gamma-ray emission,
 and 
3) only time intervals when the center of the LAT field of view is within 52$^\circ$ of the local zenith are accepted to further reduce the contamination by Earth's atmospheric emission.
We also eliminated two short periods of time during which the LAT detected the
bright GeV-emitting GRB~081024B~\citep{GRB1} and
GRB~100116A~\citep{GRB2} within 15$^\circ$ of the Cygnus Loop.
We restricted the analysis to the energy range $>200$~MeV to avoid
possible large systematics due to the rapidly varying effective area and much broader PSF at lower energies.

\section{ANALYSIS AND RESULTS}
\label{sec:ana}
\subsection{Morphological analysis}
\label{subsec:spatial}

\subsubsection{Method}

\begin{figure}
\plotone{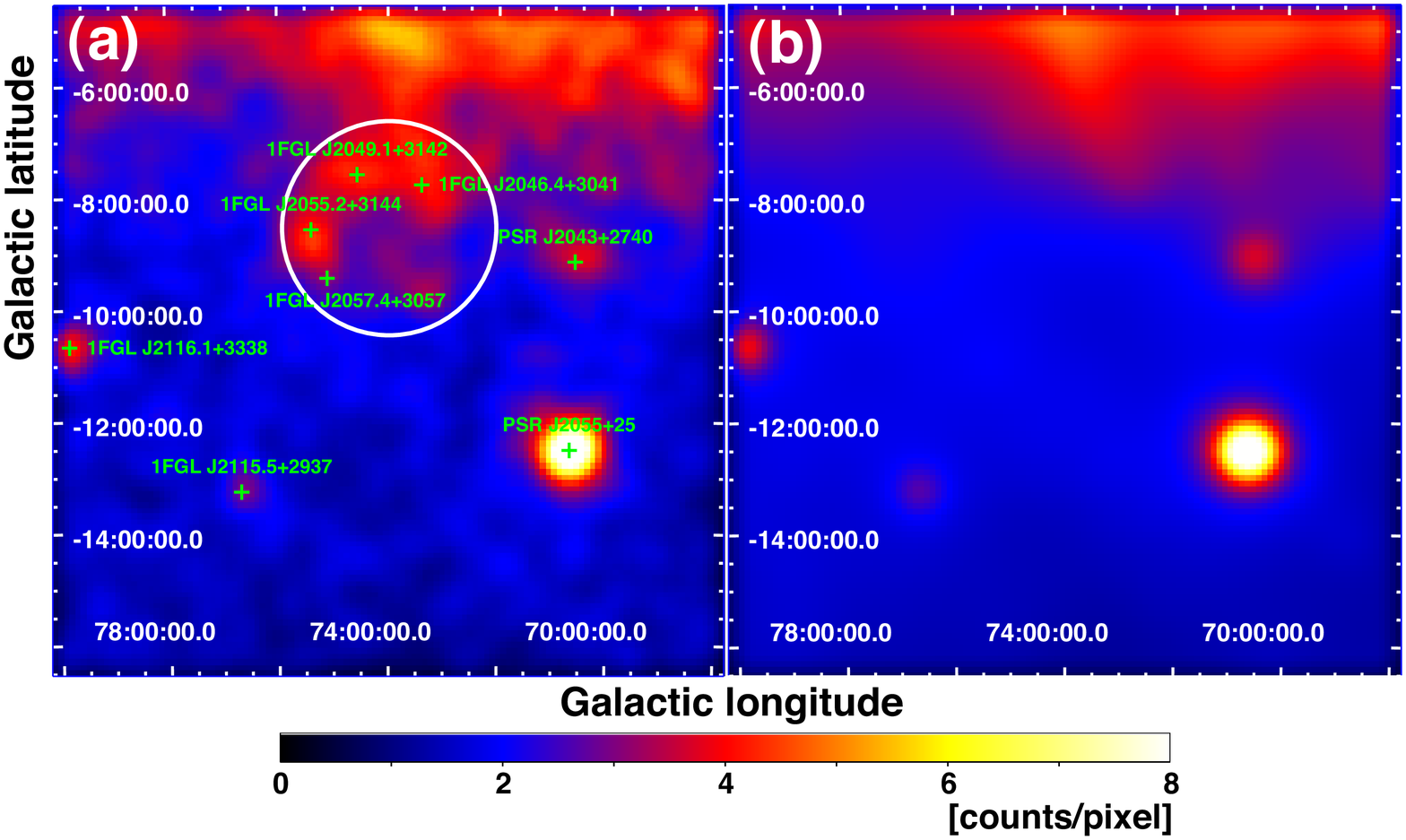}
\caption{(a)~\emph{Fermi} LAT count map in the Cygnus Loop region for photon energies 0.5--10~GeV.
The count map has a pixel size of 0.$^\circ$1 and is smoothed for display with a Gaussian kernel of $\sigma = 0.^\circ5$. Note that all along the paper the analysis is conducted on unsmoothed data taking into account the instrument PSF in the likelihood analysis.
The white circle is the location of the Cygnus Loop, defined by its radio emission.
Green crosses indicate the positions of gamma-ray sources listed in the 1FGL
catalog~\citep{1yrCatalog}.
(b)~count map expected from the background model~(taking into account the LAT PSF). The four LAT point sources associated with the Cygnus Loop are not included in the model. 
The image is binned and smoothed in the same manner as the real data.
\label{fig:raw_cmap}
}
\end{figure}

In order to study the morphology of gamma-ray emission associated with the Cygnus Loop we
performed a binned likelihood analysis based on Poisson statistics\footnote{As implemented in the publicly available
\emph{Fermi} \emph{Science Tools}. The documentation concerning the analysis tools and the likelihood fitting procedure is available from
http://fermi.gsfc.nasa.gov/ssc/data/analysis/documentation/Cicerone/.}\citep[see e.g.][]{Mattox96}.
We used only events above 0.5~GeV~(compared to the 0.2~GeV used in the spectral analysis)
 for the morphological study to take advantage of the narrower
PSF at higher energies.
For this work we used the instrument response functions~(IRFs) \texttt{P6\_V3},
which were developed following the launch to address gamma-ray detection inefficiencies that are
correlated with background rates~\citep{Rando2009}.
The analysis was performed over
a square
region of 12$^\circ \times$12$^\circ$ width with a pixel size of 0.$^\circ$1.
We set the centroid of the region to (R.A., Dec.)~$=$~(21${}^h$03${}^m$03${}^s$,
33${}^\circ$42$'$56$''$), 2$^\circ$ shifted from that of the Cygnus Loop toward negative Galactic
latitudes
to avoid the background given by Galactic diffuse emission and Galactic sources.
 Figure~\ref{fig:raw_cmap}~(a) shows a count map in the 0.5--10~GeV energy
band in the region used for the analysis, as well as the position of the Cygnus Loop from
radio measurements and point sources in the 1FGL catalog.
The four LAT sources, 1FGL~J2046.4$+$3041, 1FGL~J2049.1$+$3142, 1FGL~J2055.2$+$3144 and 1FGL~J2057.4$+$3057, are associated with the Cygnus Loop. 
Note that no gamma-ray pulsation was found for any of these LAT sources.

\subsubsection{Background model}

Although the Cygnus Loop is at intermediate Galactic latitude, the contribution of the Galactic
interstellar emission in the gamma-ray band is still important;
it must be carefully modeled to perform morphological studies. 
Some of the interstellar gas tracers
in the standard diffuse model provided by the LAT collaboration 
 are not fully adequate for the Cygnus region, notably the E(B-V) map~\citep{Schlegel98} because of infrared source contamination and temperature correction problems in such a massive-star forming region.

We therefore constructed
a dedicated diffuse emission model. The
model is analogous to the standard LAT diffuse model and it includes: a)~an isotropic background,
taking into account the isotropic diffuse gamma-ray emission as well as residual misclassified CR
interactions in the LAT; b)~large-scale Galactic inverse Compton emission produced by CR
electrons and positrons
upscattering low-energy photons, modeled using the GALPROP code \citep[e.g.][]{Porter08}; c)
emission from interstellar gas arising from nucleon-nucleon interactions and electron
Bremsstrahlung, which is modeled through spatial templates accounting for atomic gas and
CO-bright molecular gas, partitioned along the line of sight to separate the Cygnus
complex from the segments of the spiral arms in the outer Galaxy seen in this direction,  as well
as dark gas traced by visual extinction. With
respect to the standard diffuse model, this one includes higher-resolution \hi\ data
\citep{Taylor03}, visual extinction as a dark-gas tracer \citep{Rowles09,Froebrich10}
and it is specifically tuned to reproduce LAT data in the Cygnus region, including the region of the
Cygnus
Loop. All these
components, along with individual sources,
were jointly fitted to the LAT data in 10 energy bands
over the range 0.1--100 GeV with a free
normalization in each energy bin (except for the inverse Compton model that was kept fixed). For
further details we refer the reader to the dedicated paper \citep{CygnusISM}, where the model is
also discussed in detail in terms of CR and
interstellar medium properties. We note that the presence of the Cygnus Loop was taken into account
in this study. Several
models were considered for the Loop, a combination of point sources and geometric templates 
such as a disk and a ring, as in the analysis performed in this paper. In this way we verified that
the impact of the emission from the Cygnus Loop on the parameters of the global model of the region
is small \citep{luigithesis}.

The results of this analysis were used to construct two model cubes, as a function of direction and
energy, separately accounting for the isotropic and smooth large-scale Galactic inverse-Compton
emission (a and~b) and the structured emission from the gas (c). Such model cubes are part of
the background model used to study the Cygnus Loop in this paper. For each of them we included a
free normalization in order to further allow the model to adapt in the different cases we
investigated along the paper.

In addition to interstellar emission, the background model to study the Cygnus Loop includes
individual point-like sources in the 1FGL catalog within 15$^\circ$ of the Cygnus Loop except for the sources associated with the Cygnus Loop itself in the catalog; 
their positions were kept fixed at those given in the catalog and 
the spectra of the two gamma-ray pulsars in the region used for the analysis
were modeled as power laws with exponential cutoffs leaving all spectral parameters free,
while the spectra of the other sources were modeled
as power laws leaving the integral fluxes as free parameters and assuming the spectral indices reported
in the catalog.
Note that, due to the PSF, which is poor compared to other wavelengths and strongly energy dependent, and the presence of a bright and structured background given by the interstellar emission, it is difficult to mask the background sources, and they are instead modeled along with the Cygnus Loop.
The resulting model of background emission~(i.e. not including emission associated with the
Cygnus Loop) is shown in Figure~\ref{fig:raw_cmap}~(b).
The pulsars J2043$+$2740 and J2055$+$25 are the most important point sources in the vicinity, but the amount of events from those sources that fall within the Cygnus Loop~(due to the broad low-energy PSF) is only 0.4~\% and 0.2~\% of the estimated emission from the Loop, respectively.

\subsubsection{Comparison with observations at other wavelengths}

\begin{figure}
\plotone{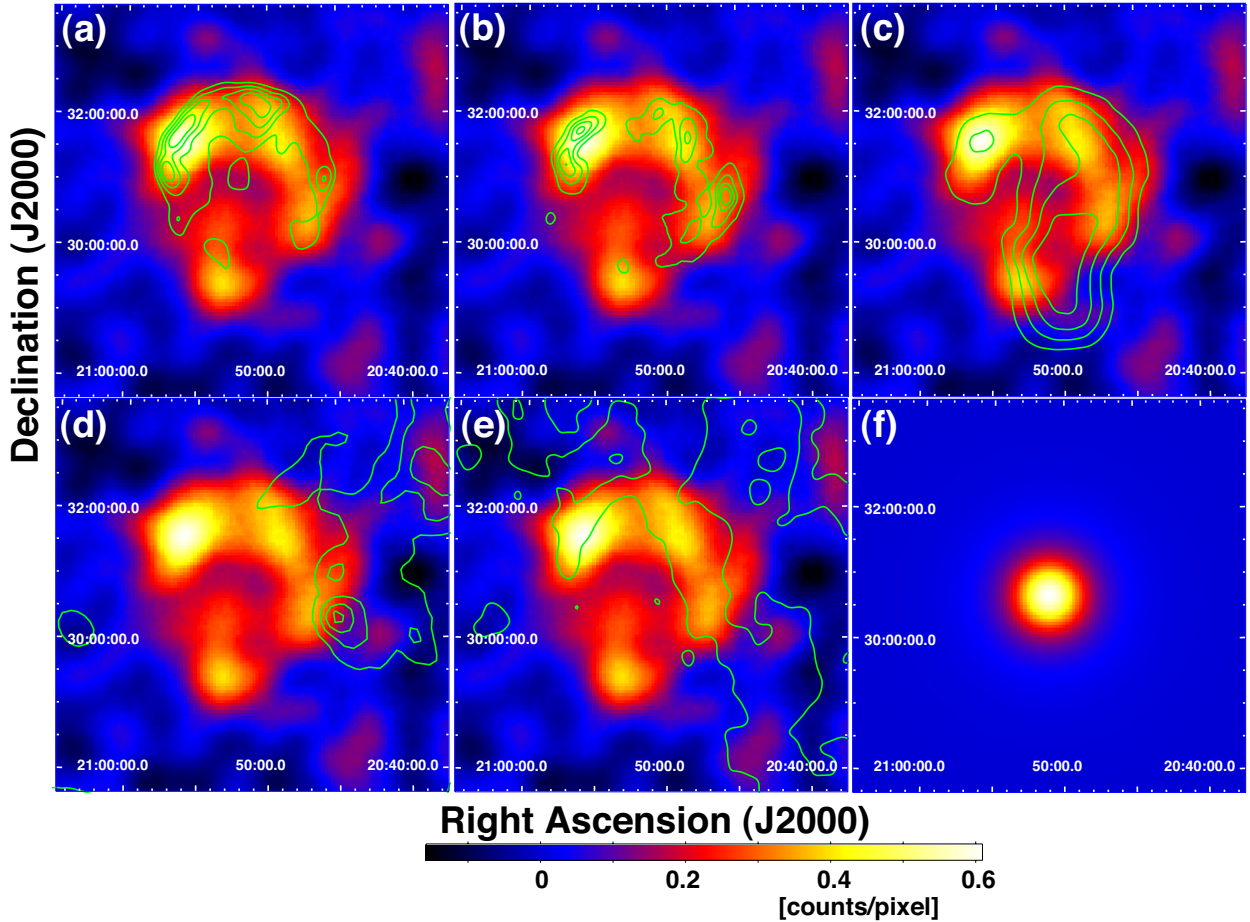}

\caption{
\footnotesize 
Background-subtracted LAT count map in the 0.5--10~GeV energy range.
The count map is binned using a grid of $0.^\circ05$ and smoothed with a Gaussian
kernel of $\sigma$ $=$
 0.$^\circ$5.
Negative residuals are shown to gauge the quality of the subtraction of the background emission.
Green contours correspond to images at different wavelengths. (a)~X-ray
count map (0.1--2~keV) by {\it ROSAT}.
Contours are at 20, 40, 60, 80~\% levels; the image was first cleaned
from background emission, estimated by 
fitting data surrounding the Cygnus Loop with a bilinear function,
and smoothed with a Gaussian kernel of $\sigma$ $=$ 0.$^\circ$2;
(b)~\halpha image obtained from the publicly available Digital Sky
Survey obtained with the same procedure explained for X-ray data.
We selected the POSS-II F~(red) filtered survey whose transmission coefficient
peaked near \halpha.
(c)~1420~MHz radio continuum emission~\citep{Bonn}; extraction of the contours as for the previous images.
(d)~ $^{12}$CO~($J=1\rightarrow0$) line intensities integrated for velocities from $-25$~km~s$^{-1}$ to 30~km~s~$^{-1}$. 
The data are taken from
the CfA survey~\citep{CO} cleaned from background using the moment-masking technique~\citep{moment_mask};
 the image was
smoothed using a Gaussian kernel with $\sigma$ $=$~0.$^\circ$25; contours are at 1, 4, 7, 10~K~km~s$^{-1}$.
(e)~The infrared intensity map at 100~$\mu$m by InfraRed Astronomical Satellite~(IRAS)~\citep{IRAS}; the image was smoothed using a Gaussian kernel of $\sigma$ $=$~0.$^\circ$2; contours are at 15, 25, 35, 45~MJy~sr$^{-1}$.
The contour at the top-right corner is the highest one.
(f)~the effective LAT PSF in the energy band of the LAT count map for a photon spectral index of 2.5.
The PSF map is binned and smoothed in the same manner as the real data.
\label{fig:subtract_cmap}}
\end{figure}

Figure~\ref{fig:subtract_cmap} shows the count map after subtracting the background
emission
 in a 6$^\circ~\times$~6$^\circ$ region centered on the Cygnus Loop,
(R.A., Dec.)~$=$~(20${}^h$51${}^m$06${}^s$, 30${}^\circ$41$'$00$''$), 
with overlays of images at different wavelengths: X-rays, \halpha line, radio continuum at
1420~MHz, 
infrared radiation at 100~$\mu$m and CO 2.6~mm line.
The correlation  between gamma rays, X-rays and \halpha emission is evident. 
There is correlation among gamma rays and radio continuum emission in the northern part of the Cygnus Loop. On the other hand,
the southern rim is brighter in radio continuum emission than in
gamma rays, a phenomenon that perhaps might be explained by the existence of another SNR overlapping with the southern
part of the Cygnus Loop~\citep{anotherSNR}.
The CO line intensities were integrated for velocities with respect to the local standard of rest $-25$~km~s$^{-1}$~$<V<30$ km~s$^{-1}$\footnote{The velocity corresponding to the distance from the Earth~(540~pc) is $\sim$~0~km~s$^{-1}$.}. 
No obvious association with CO emission is found; 
some molecular material is apparently located on the Western side of the Cygnus Loop, 
but the relationship is not clear. 
On the other hand, some correlation with thermal emission from dust at 100~$\mu$m, which can be considered as a proxy of total interstellar matter densities, is possible.


To quantitatively evaluate the correlation with emission at other wavebands, 
we fitted the LAT counts with the different models for the Cygnus Loop on top of the background model described above.
First the Cygnus Loop was modeled with the four 1FGL sources, and then using the images at other wavelengths as spatial templates assuming a simple power-law spectrum.
Note that we did not use the CO and infrared images as spatial templates due to clear differences between them and the gamma-ray image as shown in Figure~\ref{fig:subtract_cmap}.
 The resulting maximum likelihood values with respect to the maximum likelihood for the null
hypothesis (no emission associated with the Cygnus Loop) are summarized in
Table~\ref{tab:likeratio}.
The test statistic~(TS) values, i.e. $-2\ln({\rm likelihood~ratio})$~\citep[e.g.][]{Mattox96}, for the X-ray and \halpha images are significantly larger 
than for the four 1FGL sources.
On the other hand, the TS for the radio image
increases moderately 
 in spite of the association in the northern rim, confirming that radio continuum structures in the southern rim do not well correlate with gamma-ray emission.

\subsubsection{Geometrical Models}
\begin{figure}
\plotone{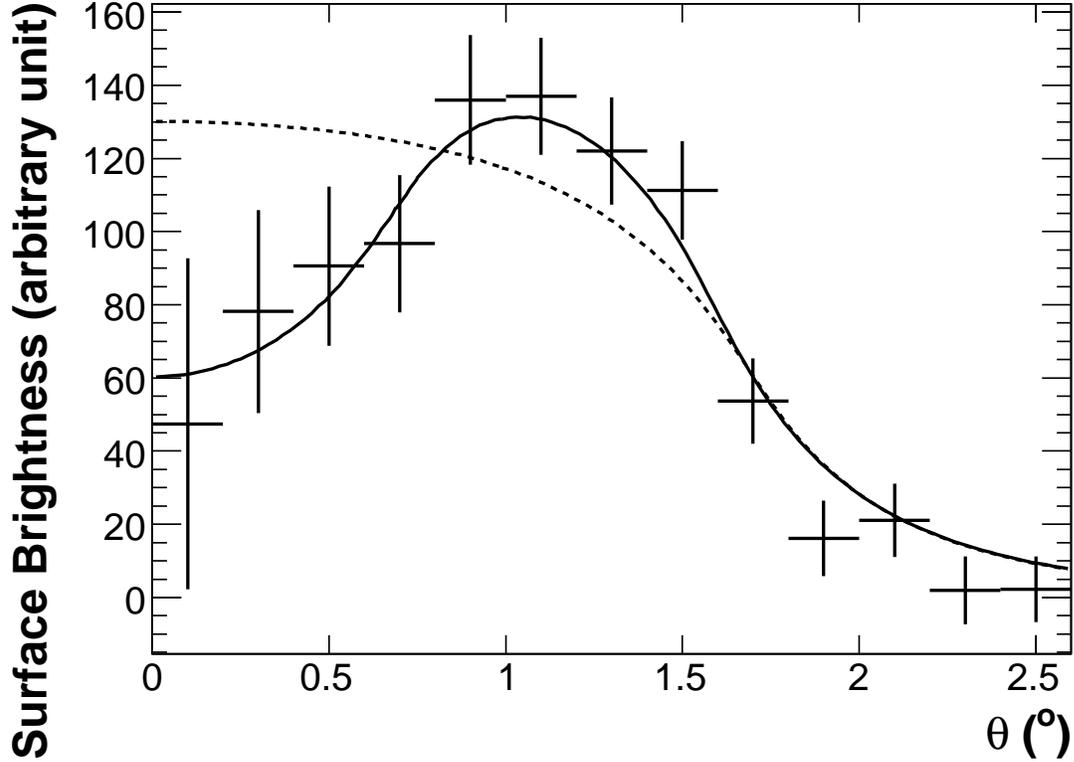}
\caption{Radial profile of the Cygnus Loop in gamma rays in the 0.5--10~GeV energy range~(crosses). 
The origin is the center of the best-fit ring model.
Gamma-ray data have the background emission subtracted.
Note that the data are not smoothed.
Overlaid are the distributions expected for the best-fit ring shape~(solid line) and the best-fit
disk shape~(dotted line) as emission surfaces, with parameters fit to gamma-ray data taking the
LAT instrument response into account. Details of the fits are described in the text.
 \label{fig:radialProfile}}
\end{figure}

We further characterized the morphology of gamma-ray emission associated with the Loop by using simple parametrized geometrical models.
We started with a uniform disk/ring
assuming a simple power-law spectrum.
We varied the radius and location of the disk 
and evaluated the maximum likelihood values.
In the case of the ring, we varied inner and outer radii as well.
The resulting TS values are reported in Table~\ref{tab:likeratio}.
The TS value for the ring with respect to the disk shape is
$\simeq$~12.
Assuming that, in the null hypothesis, the TS value is distributed as a $\chi^2$ with $n$ degrees of freedom, where $n$ is the difference in degrees of freedom between the two nested models compared
\footnote{see link to \emph{Fermi} \emph{Science Tools} Cicerone} ($n=1$ in the present case), it would be equivalent to an improvement at $\sim~3.5~\sigma$ confidence level. Let us note, however, that the conversion of TS values into confidence level~(or, equivalently, false positive rate) is subject to numerous caveats, see e.g. \citet{TSconversion}. We will thus take into account the source morphology uncertainties in the spectral analysis, below.
In order to further illustrate the morphology of the gamma-ray emission, in
Figure~\ref{fig:radialProfile} we show its radial profile compared with the best-fit disk/ring models. 

\begin{figure}
\plotone{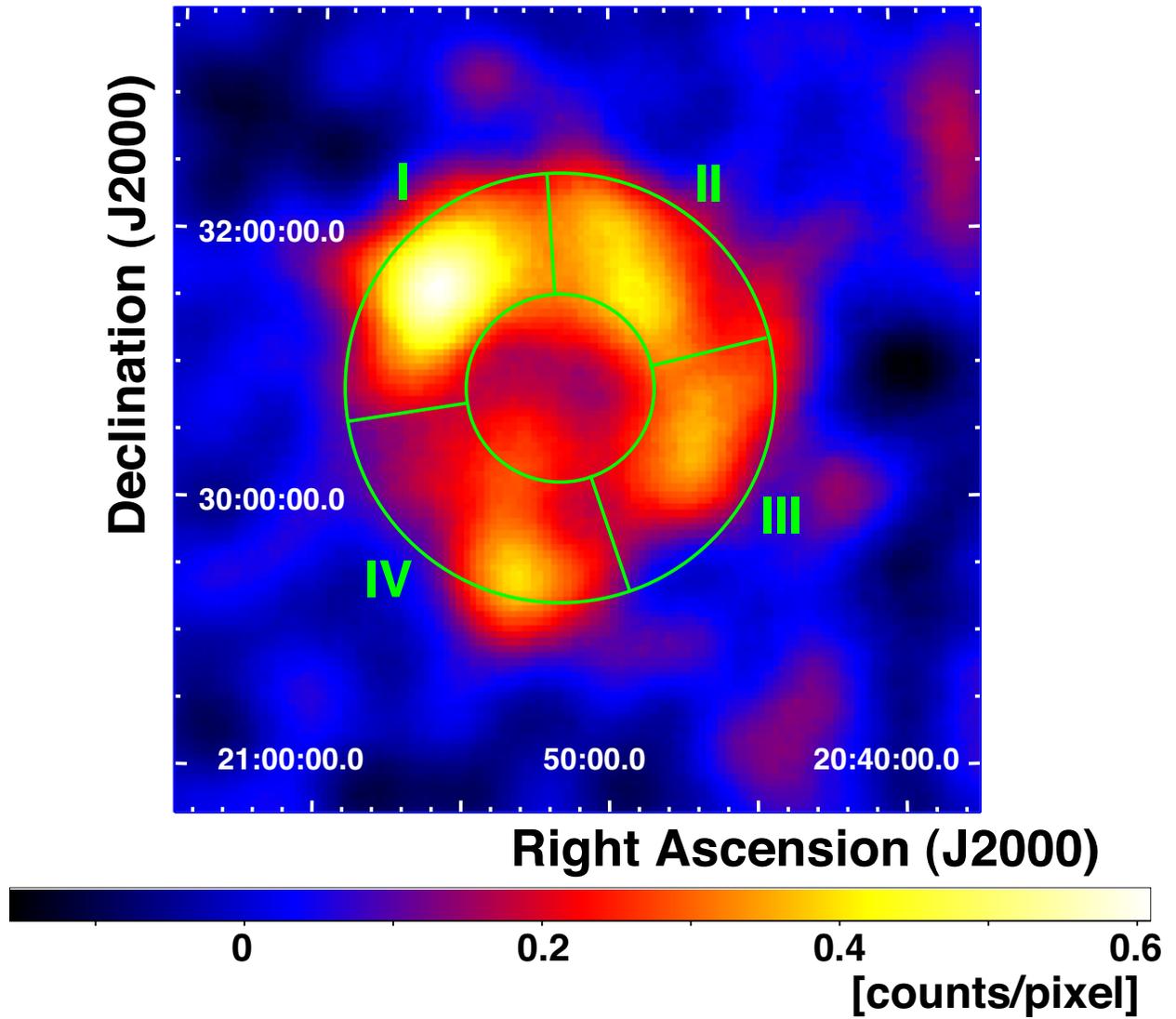}
\caption{Definitions of the regions
 of the Cygnus Loop used for the morphology analysis
(\S~\ref{subsec:spatial}) overlaid on the LAT count map as shown in Figure~\ref{fig:subtract_cmap}.
 \label{fig:rim_definition}}
\end{figure}

Finally, we want to verify if there are any spectral variations in the gamma-ray emission
associated with the Cygnus Loop we are modeling as a whole. We thus divided
the best-fit ring into four regions
as shown in Figure~\ref{fig:rim_definition} and allowed an
independent
normalization and spectral index for the four portions of the ring.
There was no significant improvement of the likelihood for such a non-uniform ring.
The TS value and power-law spectral index for each of the
four regions 
 of the remnant are
reported for
reference in Table~\ref{tab:spatial_difference}. No significant differences are found between the
four spectral indexes.
Therefore we adopted the uniform ring template with maximum likelihood parameters
for the whole
SNR in the following spectral analysis.

\subsection{Spectral analysis} 

\begin{figure}
\plotone{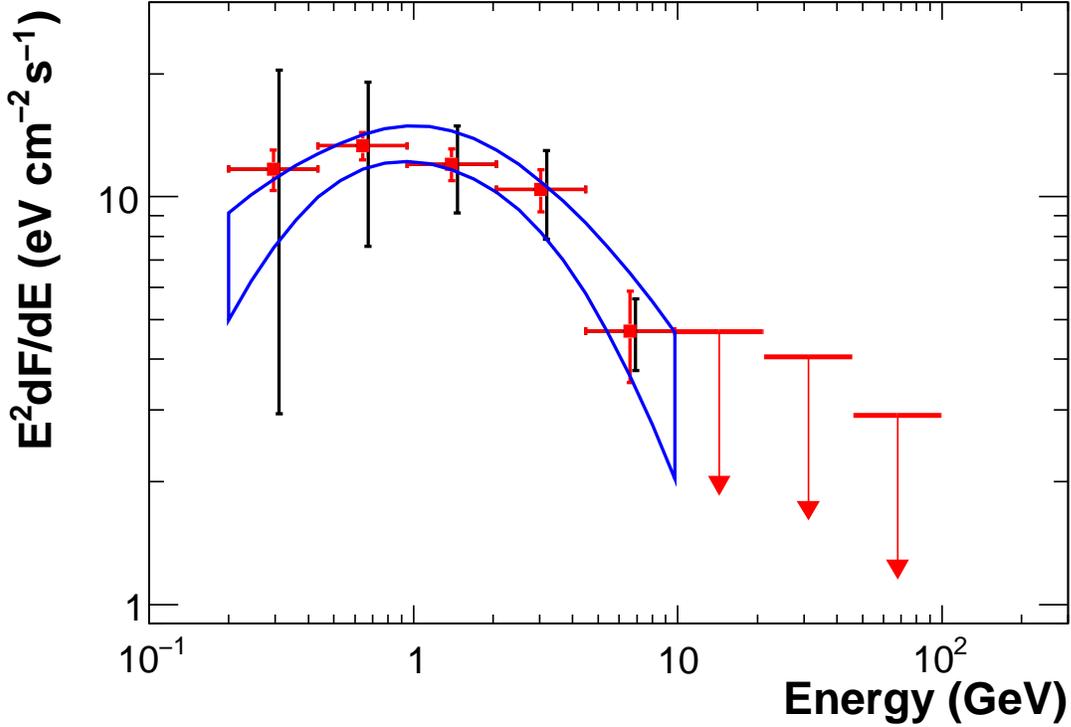}
\caption{ Spectral energy distribution of the gamma-ray emission measured by the LAT for the
Cygnus Loop. 
Red squares are LAT flux points.
Horizontal bars indicate the energy range the flux refers to.
Vertical bars show statistical errors in red and systematic errors~(added in quadrature for illustration purposes) in black. 
In energy bins where
the detection is not significant (test statistic $<$~10) we show upper limits at the
90~\% confidence level.
 The blue region is the 68~\% confidence range~(no systematic error) of the LAT spectrum assuming that the spectral shape is a log parabola.
 \label{fig:spec}}
\end{figure}

To measure the spectrum we made maximum likelihood fits in 8 logarithmically-spaced energy bands from 0.2~GeV to 100~GeV, using the ring template as the model for the spatial distribution of the Cygnus Loop. 
Figure~\ref{fig:spec} shows the resulting spectral energy distribution (SED).
Upper limits at the 90~\% confidence level are calculated assuming a photon
index of 2 if the detection is not significant in an energy bin, i.e., the TS value with respect to the null hypothesis is less than 10~(corresponding to 3.2~$\sigma$
for one additional degree of freedom).
Note that the value of the spectral index has a negligible effect on the upper limits.

We identify at least three different sources of systematic uncertainties affecting the
estimate of the fluxes: uncertainties in the LAT event selection efficiency, the morphological
template and the diffuse model adopted for analysis.
Uncertainties in the LAT effective area are estimated to be 10~\% at 100~MeV, decreasing to 5~\% at
500~MeV, and increasing to 20~\% at 10~GeV and above~\citep{Rando2009}.
Evaluating the systematic uncertainties due to the modeling of interstellar emission is a challenging task, because interstellar emission is highly structured and methods used at other wavelengths, like comparisons with neighboring regions, are not fully adequate in the GeV band. We therefore roughly gauged the related uncertainties by comparing the results with those obtained 
by adopting instead the standard LAT diffuse background models
\footnote{\emph{gll\_iem\_v02} and \emph{isotropic\_iem\_v02} available from the FSSC\\
http://fermi.gsfc.nasa.gov/ssc/data/access/lat/BackgroundModels.html}. 
We similarly gauged the uncertainties due to the morphological template by comparing the results with those obtained by using the best-fit disk template instead of the ring.
The total systematic errors are set by adding the above uncertainties in quadrature.
Systematic uncertainties are driven by the imperfect knowledge of the background
emission and, especially below a few hundred MeV, of the LAT response.
In Figure~\ref{fig:spec} we show the uncertainties obtained following these prescriptions.

We probed for a spectral break in the LAT energy
band by comparing the likelihood values of a spectral fit over the whole energy range considered 
based on a simple power law and other spectral functions.
Note that no systematic uncertainties are accounted for in the likelihood fitting process.
The TS values and best-fit parameters are summarized in Table~\ref{tab:spectral_shape}.
The fit with a log-parabola function yields a TS value of $\sim 50$ compared to a simple power-law model, which corresponds to an improvement at the $\sim 7$~$\sigma$ confidence level.
In spite of the uncertainties discussed above in the estimate of the confidence level, the large TS value is indicative of a significant improvement in the fit.
A smoothly broken power law provides a very slight increase in the likelihood with respect to the log-parabola function, while a power law with exponential cutoff gives a worse fit.
In conclusion, a simple power law as spectral model can be significantly rejected and
from all the different models with cutoffs we get evidence for a steepening of the spectrum
above 2--3~GeV.
We detect gamma-ray emission with a formal significance of 23~$\sigma$
for the above curved spectral shapes.
The observed photon flux and energy flux in the 0.2--100~GeV range are $5.0^{+0.6}_{-0.6} \times 10^{-8}$~cm$^{-2}$~s$^{-1}$ and $6.5^{+0.7}_{-0.6} \times 10^{-11}$~erg~cm$^{-2}$~s$^{-1}$, respectively.


\section{DISCUSSION}
\label{sec:discuss}
The gamma-ray luminosity inferred from our analysis is $\sim$~$1 \times 10^{33}$erg~s$^{-1}$
between 1--100~GeV, lower by one order of magnitude than observed for other GeV-emitting
SNRs~\citep[typically $> 10^{34}$~erg~s$^{-1}$,][]{LAT-W51C, LAT-W44, LAT-IC443, LAT-W28, LAT-W49B}.
The spatial distribution is best represented by a ring with inner/outer radii 0$^\circ$.7~$\pm$~0$^\circ$.1 and 1$^\circ$.6~$\pm$~0$^\circ$.1,
respectively. 
This makes the Cygnus Loop the largest gamma-ray emitting SNR observed so far, allowing
us to perform a detailed morphological comparison with emission at other wavelengths.

There is a correspondence
 among gamma-ray emission, 
 X-ray rims
 and
\halpha filaments, indicating that the high-energy particles responsible for gamma-ray emission
are in the vicinity of the shock regions.
The Balmer-dominated filaments define the current location of the blast wave and  mark the presence of neutral material. Detailed studies of the particular locations at the northeast have used these nonradiative shocks as density probes~\citep{density1,density2,density3} and derived post-shock densities of $\sim$~5~cm$^{-3}$ where gamma-ray emission is expected to be bright due to the compressed material and high density of accelerated particles.

The radio continuum emission, originated by high-energy electrons via synchrotron radiation,
is well correlated with gamma-ray emission in the northern region of the remnant but not in the
southern one 
The presence of a second SNR was suggested by \cite{anotherSNR}.
The two SNRs would be at about the same distance based on the rotation measure analysis of the radio data~\citep{anotherSNRdistance}.
The lack of correlation between gamma rays and radio continuum emission in the southern region
plausibly implies that the second SNR is not producing significant gamma-ray emission at our current sensitivity.
There might be some correlation between total matter densities as traced by infrared thermal emission from dust and gamma-ray emission, whereas CO emission does not obviously overlap with the Cygnus Loop.

From these considerations,
 we argue that the bulk of gamma-ray emission comes from interactions of high-energy particles
accelerated at the shocks of the Cygnus Loop with interstellar matter or fields in the regions
just adjacent to the shocks with a gas density of $\sim$~5~cm~$^{-3}$.

To model the broadband emission from the entire SNR we adopt the simplest possible assumption
that gamma rays are emitted by a population of accelerated protons and electrons distributed in the same region 
and characterized by constant matter density and magnetic field strength. 
We assume the injected electrons to have the same momentum distribution as protons.
This assumption requires a break in the momentum spectrum
because the spectral index in the radio domain, corresponding to lower
particle momenta, is much harder than for gamma rays, which
correspond to higher particle momenta.
Therefore, we use the following functional form to model the momentum distribution of injected
particles:
\begin{equation}
Q_{e, p}(p) = a_{e,p} \left( \frac{p}{1~{\rm GeV}~c^{-1}} \right)^{-s_{\rm L}} \left\{ 1+\left(\frac{p}{p_{\rm br}} \right)^2 \right\} ^{-(s_{\rm H}-s_{\rm L})/2},
\end{equation}
 where $p_{\rm br}$ is the break momentum, $s_{\rm L}$ is the spectral index below the break
and $s_{\rm H}$ above the break.
Note that here we consider minimum momenta
of 100~MeV~$c^{-1}$ since the details of the proton/electron injection process are poorly known.

Electrons suffer energy losses due to ionization, Coulomb scattering, Bremsstrahlung,
synchrotron
emission and inverse Compton~(IC) scattering.
We calculated the evolution of the electron momenta spectrum by the following equation:
\begin{equation}
  \frac{\partial N_{e,p}}{\partial t} = \frac{\partial}{\partial p} \left( b_{e,p}N_{e,p} \right) + Q_{e,p} ,
\end{equation}
where $b_{e,p} = -dp/dt$ is the momentum loss rate, and $Q_{e,p}$ is the particle injection rate.
We assume $Q_{e,p}$ to be constant, i.e., that the  
shock produces a constant number of particles 
until the SNR enters the radiative phase, at which time the source turns off. 
This prescription approximates the weakening of the shock and the reduction in the particle acceleration efficiency, which would be properly treated by using a time-dependent shock compression ratio~\citep{SNRaccelerationTime}. 
To derive the remnant emission spectrum we calculated $N_{e,p}(p,T_{\rm 0})$
numerically, where $T_{\rm 0}$ is the SNR age of $2 \times 10^4$~yr.
Note that we neglected the momentum losses for protons since the timescale of neutral pion
production is $\sim$~10$^7/\nh$~yr where $\nh$ is the gas density averaged over the entire SNR
shell and is much longer than the SNR age.
Also we do not consider the gamma-ray emission by secondary positrons and electrons from
charged pion decay, 
because the emission from secondaries is generally unimportant relative to that from primary
electrons 
unless the gas density is as high as that in dense molecular clouds 
and the SNR evolution reaches the later stages,
or the injected electron-to-proton ratio
is much lower than locally observed.
The gamma-ray spectrum from $\pi^0$ decay produced by the interactions of protons with
ambient hydrogen is calculated based on \cite{Dermer86}
using a scaling factor of 1.84 to account for helium and heavier nuclei in target
material and cosmic rays~\citep{Mori09}. 
Contributions from bremsstrahlung and inverse Compton scattering by accelerated electrons
are computed based on \cite{Blumenthal70}, whereas synchrotron radiation is based on \cite{Crusius86}.

\begin{figure}
\plotone{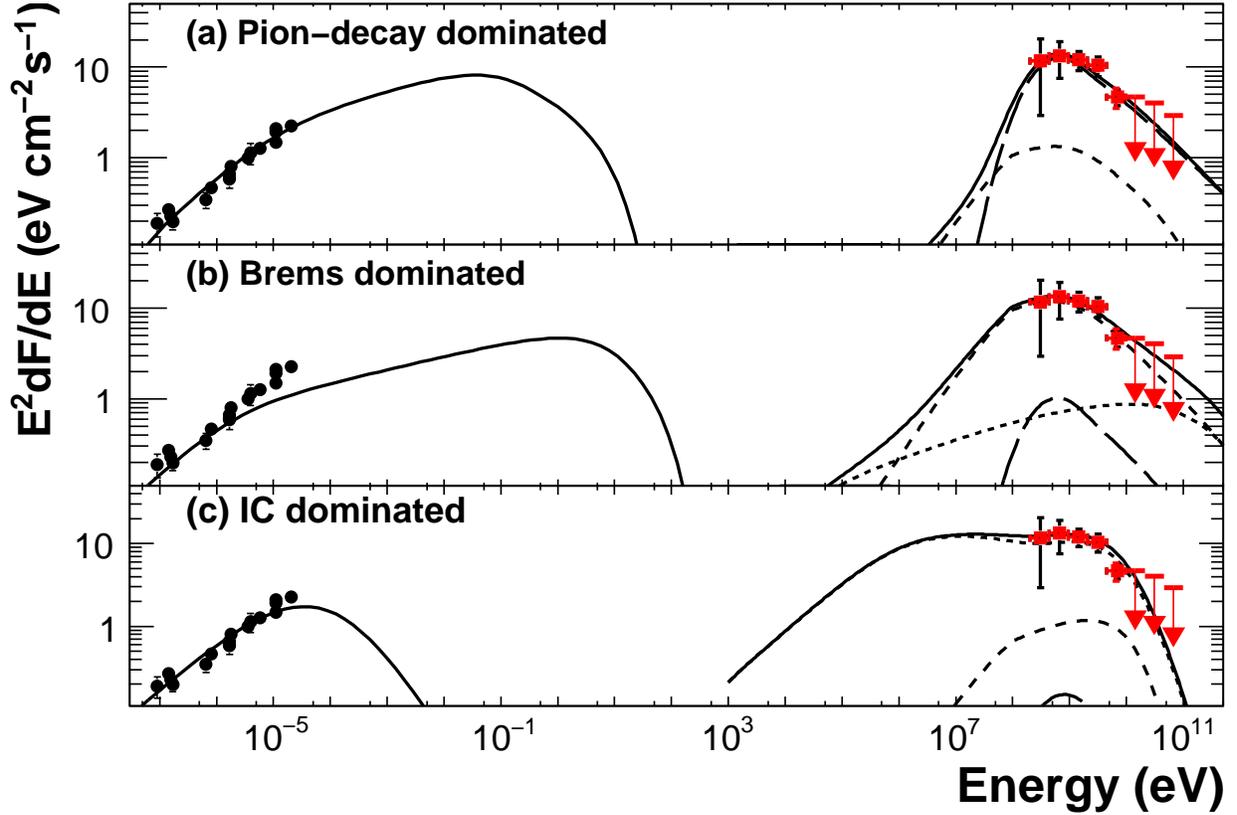}
\caption{Multi-band spectrum of the Cygnus Loop.
 \label{fig:spec_multi}
In the GeV band LAT measurements are reported as in Figure~\ref{fig:spec}.
The radio continuum emission~\citep{RadioSpectrum} is shown by black dots. Radio emission is
modeled as synchrotron
radiation, 
while gamma-ray emission is modeled by different combinations of $\pi^0$-decay~(long-dashed
curve),
bremsstrahlung~(dashed curve), and inverse Compton~(IC) scattering~(dotted
 curve). 
Details of the models are described in the text: a) $\pi^0$-decay dominated model, b)
bremsstrahlung dominated model, c) IC dominated model.
}
\end{figure}

First, we consider a $\pi^0$-decay dominated model.
The number index of protons in the high-energy regime
 is constrained to be $s_{\rm H} \approx 2.6$ from the gamma-ray spectral
 slope.
The spectral index of the proton momentum below the break is determined to be
$s_{\rm L} \approx 1.8$ by modeling the radio spectrum as synchrotron radiation by relativistic
electrons (under the assumption that protons and electrons have identical injection spectra). The spectral index
$\alpha$ of the radio continuum emission is $\sim$~0.4~\citep{RadioSpectrum}, where $\alpha$ is defined as $S_\nu \propto \nu^{-\alpha}$ with $S_\nu$ and $\nu$ the flux density and the frequency, respectively.
It is difficult to derive the break momentum of the proton spectrum from the gamma-ray
spectrum, since in the GeV energy band we expect a curvature due to
kinematics of $\pi^0$ production and decay.
The gamma-ray spectrum provides thus only an upper bound for the momentum break at
$\sim$~10~GeV~$c^{-1}$.
On the other hand, the momentum break cannot be lower than $\sim$~1~GeV~$c^{-1}$ to avoid conflicts
with radio data.
We adopt a break at the best-fit value, 2~GeV~$c^{-1}$.
The resulting total proton energy,  $W_{p}\sim 2.6\times10^{48}\cdot(5\
\mathrm{cm}^{-3}/\nh)\cdot(d/540\ \mathrm{pc})^{2}$~erg, is less than 1~\% 
of the typical kinetic energy of a supernova explosion.
For an electron-to-proton ratio $K_{ep} = 0.01$ at 1~GeV~$c^{-1}$, which
is the ratio measured at the Earth,
the magnetic field strength is constrained to be $B \sim 60\;\upmu$G by radio data.
The magnetic field strength of the undisturbed medium in the northeastern rim was estimated to
be $\sim$~20~$\upmu$G by \cite{undisturbed_magnetic_field}
based on the measurements of shell thickness and expansion velocities together with the theory of hydromagnetic shock propagation 
given the density of the undisturbed medium $\sim$~1~cm$^{-3}$~\citep[e.g.,][]{density3}.
The compression behind the shock front 
indicates a magnetic field strength
similar to
the value used above in the modeling. 
Using the parameters summarized in Table~\ref{tab:model}, 
we obtained the SEDs shown in Figure~\ref{fig:spec_multi}~(a).

It is difficult to model the gamma-ray spectrum with a model dominated by electron bremsstrahlung
because the break in the electron spectrum required to reproduce the gamma-ray spectrum
would appear in the radio domain as shown in Figure~\ref{fig:spec_multi}~(b).

The gamma-ray spectrum can be reproduced by an inverse Compton dominated model shown in
Figure\ref{fig:spec_multi}~(c).
Gamma-ray emission of IC origin is
 due to interactions of high-energy electrons with optical and infrared radiation fields and the cosmic microwave background~(CMB).
We used in our calculations the first two components as they are modeled in~\citet{Porter08} at the location of the Cygnus Loop.
Since their spectra are very complex, they are approximated by two
infrared and two optical blackbody components.
The flux ratio between the IC and the synchrotron components 
constrains the magnetic field to be less than 2~$\upmu$G
and requires a low gas density of $\nh \sim 2 \times 10^{-2}~\mathrm{cm}^{-3}$ to suppress the
electron bremsstrahlung.
Although such a low density may exist inside the remnant based on X-ray
observations~\citep[e.g.,][]{EinsteinXray}, 
gamma-ray emission peaks at the shock regions where the gas density is
$\sim$~$1-5$~cm~$^{-3}$ (see above).
Increasing the intensity of the interstellar radiation field
would loosen the constraint on the
gas density. However, a radiation field about 50 times more intense is required to
satisfy the above assumption on the gas density. 

To summarize, it is most natural to assume that gamma-ray emission from the Cygnus Loop is dominated by decay of $\pi^0$ produced in nucleon-nucleon interactions of hadronic cosmic rays with interstellar matter.
It should be emphasized that
our observations of the Cygnus Loop combined with the radio data constrain the proton momentum break to be in the range, 1--10~GeV~$c^{-1}$,
despite the lack of association with dense molecular clouds unlike the other middle-aged SNRs detected with the LAT.
Thus in this case cosmic rays responsible for gamma-ray emission are localized near their acceleration sites without significant diffusion taking place.
The correspondence observed between gamma rays and \halpha emission may be accounted for in the ``crushed cloud'' scenario by \citet{UchiyamaModel},
 although the expected filaments cannot be resolved by current gamma-ray telescopes.
The predictions by \cite{twostep_acceleration}
cannot be directly compared to the Cygnus Loop since 
their simulations were performed for environments characterized by dense clouds.
However, the scenario of acceleration by reflected shocks 
 might be operative, on consideration of
X-ray and
optical observations~\citep[e.g.,][]{reflected_shock}.

\section{CONCLUSIONS}
\label{sec:conclusion}

We analyzed gamma-ray measurements by the LAT in the region of the Cygnus Loop, detecting
significant gamma-ray emission associated with the remnant. 
The gamma-ray luminosity is $\sim$~$1 \times 10^{33}$erg~s$^{-1}$ between 1--100~GeV,
lower than for other GeV-emitting SNRs studied with LAT data.
The morphology of gamma-ray emission is best represented by a ring with inner/outer
radii 0.$^\circ$7~$\pm$~0$^\circ$.1 and 1.$^\circ$6~$\pm$~0$^\circ$.1.
The Cygnus Loop is thus the most extended gamma-ray emitting SNR detected in the GeV~band so far
and the morphology of gamma-ray emission can be compared in detail with observations at other
wavelengths.
There is correspondence among gamma rays, the X-ray rims and the \halpha
filaments, 
indicating that the high-energy particles responsible for the gamma-ray emission are in the vicinity of the shock regions.

The gamma-ray spectrum has a break in the 2--3~GeV energy range.
The decay of $\pi^0$ produced by interactions of hadrons accelerated
by the remnant with interstellar gas naturally explains the gamma-ray spectrum.
In this scenario our observations of the Cygnus Loop indicate that the proton momentum
spectrum 
is steep in the high-energy regime,
with a spectral break which is constrained together with radio continuum emission in the range
1--10~GeV~$c^{-1}$.
The absence of molecular clouds in the areas of gamma-ray emission~(contrary to other middle-aged
{\emph Fermi} SNRs) constrains some of the scenarios invoked to explain the observed spectral properties of GeV emitting SNRs.


\acknowledgments
The \textit{Fermi} LAT Collaboration acknowledges generous ongoing support
from a number of agencies and institutes that have supported both the
development and the operation of the LAT as well as scientific data analysis.
These include the National Aeronautics and Space Administration and the
Department of Energy in the United States, the Commissariat \`a l'Energie Atomique
and the Centre National de la Recherche Scientifique / Institut National de Physique
Nucl\'eaire et de Physique des Particules in France, the Agenzia Spaziale Italiana
and the Istituto Nazionale di Fisica Nucleare in Italy, the Ministry of Education,
Culture, Sports, Science and Technology (MEXT), High Energy Accelerator Research
Organization (KEK) and Japan Aerospace Exploration Agency (JAXA) in Japan, and
the K.~A.~Wallenberg Foundation, the Swedish Research Council and the
Swedish National Space Board in Sweden.

Additional support for science analysis during the operations phase is gratefully
acknowledged from the Istituto Nazionale di Astrofisica in Italy and the Centre National d'\'Etudes Spatiales in France.

We have made use of the {\it ROSAT} Data Archive of the Max-Planck-Institut f$\ddot{\rm u}$r extraterrestrische Physik (MPE) at Garching, Germany.
The Digitized Sky Surveys were produced at the Space Telescope Science Institute under U.S. Government grant NAG W-2166. The images of these surveys are based on photographic data obtained using the Oschin Schmidt Telescope on Palomar Mountain and the UK Schmidt Telescope. The plates were processed into the present compressed digital form with the permission of these institutions.

\begin{table}
\begin{center}
\caption{ Test Statistics for Different Spatial Models Compared with the Null Hypothesis of No
Gamma-ray Emission Associated with the Cygnus Loop~(0.5--100~GeV) \label{tab:likeratio}}
\begin{tabular}{lccc}
\tableline\tableline
 Model   & Test Statistic\tablenotemark{a} &  Additional Degrees of Freedom  \\
\tableline
   Null hypothesis\tablenotemark{b}   &       0 &  0 \\
   4 point sources\tablenotemark{c} &  318 &  8 \\
   {\it ROSAT} X-rays~(0.1--2keV)\tablenotemark{d} & 406 & 2\\
   \halpha\tablenotemark{d} & 434 & 2 \\
   1420MHz radio continuum\tablenotemark{d} &  343  & 2 \\
  Uniform disk\tablenotemark{e} & 441  & 5 \\
 Uniform ring\tablenotemark{f} &   453  & 6 \\
 Non-uniform ring\tablenotemark{g} &  464  & 12 \\
\tableline
\end{tabular}
\tablenotetext{\rm a}{$-2\ln (L_{\rm 0}/L)$,
 where $L$ and $L_{\rm 0}$ are the maximum likelihoods for the model with/without the source component, respectively.}
\tablenotetext{\rm b}{Background only (no model for the Cygnus Loop).}
\tablenotetext{\rm c}{The four sources listed in the 1FGL source list associated
with the Cygnus Loop~\citep{1yrCatalog}.}
\tablenotetext{\rm d}{Background-subtracted as described in
Figure~\ref{fig:subtract_cmap}.}
\tablenotetext{\rm e}{The best fit parameters are: radius 1$^\circ$.7~$\pm$~0$^\circ$.1 and centroid (R.A.,
Dec.)~$=$~(20${}^h$52${}^m$, 30${}^\circ$50$'$).
 The error of the centroid is 0$^\circ$.04 at 68~\% confidence level. }
\tablenotetext{\rm f}{The best fit parameters are: inner/outer radii 0$^\circ$.7~$\pm$~0$^\circ$.1,
and 1$^\circ$.6~$\pm$~0$^\circ$.1, centroid (R.A.,
Dec.)~$=$~(20${}^h$51${}^m$, 30${}^\circ$50$'$). 
The error of the centroid is 0$^\circ$.04 at 68~\% confidence level.}
\tablenotetext{\rm g}{The best-fit ring were divided into four regions as shown in Figure~\ref{fig:rim_definition} and allowed an independent normalization and spectral index for the four portions of the ring.}
\end{center}
\end{table}

\begin{table}
\begin{center}

\caption{Test Statistics and Power-law Spectral Indexes for the Four Regions
 of the Remnant
as Defined in Figure~\ref{fig:rim_definition}~(0.5--100~GeV)
 \label{tab:spatial_difference}}
\begin{tabular}{lcc}
\tableline\tableline
 Region & Test Statistic\tablenotemark{a}   &  Spectral Index  \\
\tableline
 {I} & 143  &  2.49 $\pm$ 0.10  \\
 {II} & 73 & 2.32 $\pm$ 0.12  \\
 {III} & 64  & 2.25 $\pm$ 0.15  \\ 
 {IV} & 41  & 2.37 $\pm$ 0.14 \\
\tableline
\end{tabular}

\tablenotetext{\rm a}{$-2\ln (L_{\rm 0}/L)$,
 where $L$ and $L_{\rm 0}$ are the maximum likelihoods for the model with/without
the source component, respectively.}

\end{center}
\end{table}

\begin{table}
\begin{center}
\caption{Test Statistics and Parameters for Various Spectral Models~(0.2--100~GeV) \label{tab:spectral_shape}}
\begin{tabular}{lcccc}
\tableline\tableline
 Spectral Model  & Test Statistic\tablenotemark{a}  &  Degrees  & Spectral Parameters  \\
  &  &   of Freedom & \\
\tableline\tableline
Power law  & 0  & 2 & $E^{-p}$; $p=2.23\pm0.02$ \\
\tableline
Power law with  &  42 & 3 & $E^{-p}\exp{\left(-\frac{E}{E_{\rm b}}\right)}$; \\
exponential cutoff &   &  &  $p=1.57 \pm 0.12$   \\ 
 & & &   $E_{\rm b}=3.02 \pm 0.65$~GeV \\
\tableline
Log Parabola  & 50 &  3 &   $\left(\frac{E}{1~{\rm GeV}}\right)^{-p_1-p_2\log{\left(\frac{E}{1~{\rm
GeV}}\right)}}$ \\
 & & &  $p_1=2.02 \pm 0.03$ \\
 & & &  $p_2=0.27 \pm 0.02$ \\
\tableline
Smoothly broken power law  & 51  & 4  & $E^{-p_1}\left\{ 1+\left(\frac{E}{E_{\rm b}}\right)^{\frac{-p1+p2}{0.2}} \right\}^{-0.2}$    \\
 & & &  $p_1=1.83 \pm 0.06$ \\
 & & &  $p_2=3.23 \pm 0.19$ \\
 & & &  $E_{\rm b}=2.39 \pm 0.26$~GeV \\
\tableline\tableline
\end{tabular}
\tablenotetext{\rm a}{$-2\ln (L_{\rm 0}/L)$,
 where $L$ and $L_{\rm 0}$ are the maximum likelihood values for the model under consideration and
the power-law model, respectively.}
\tablecomments{The test statistics for the best-fit uniform ring with exponential cutoff, log parabola, and smoothly broken power law with respect to the null hypothesis of no emission associated with the Cygnus Loop are 572, 580, and 581 in the energy band 0.2--100~GeV.
}
\end{center}
\end{table}

\begin{table}
\begin{center}
\caption{Model parameters for the Cygnus Loop.\label{tab:model}}
\begin{tabular}{lccccccccc}
\tableline\tableline
 Model  & \kep\tablenotemark{a} &  $s_{\rm L}$\tablenotemark{b} & $p_{\rm br}$\tablenotemark{c} & $s_{\rm H}$\tablenotemark{d} & $B$ &
 $\nh$\tablenotemark{e} & $W_{p}$\tablenotemark{f} & $W_{e}$\tablenotemark{f} \\
   &    &  & (GeV~$c^{-1}$) & &  ($\upmu$G) & (cm$^{-3}$) & ($10^{48}$~erg) & ($10^{48}$~erg) \\
\tableline
(a)~Pion & 0.01  & 1.8 & 2 & 2.6 & 60 & 5 & 2.6 & 4.9~$\times$~$10^{-2}$ \\
(b)~Bremsstrahlung & 1 & 1.8 & 2 & 2.7 & 12 & 5  & 0.21 & 0.43  \\
(c)~Inverse Compton\tablenotemark{g} & 1 & 1.8 & 25 & 5.0 & 1.8 & 0.02 & 5.9 & 9.8 \\
\tableline
\end{tabular}
\tablenotetext{\rm a}{The ratio electrons-to-protons
at 1~GeV~$c^{-1}$.}
\tablenotetext{\rm b}{
The momentum distribution of particles
 is assumed to be a smoothly broken power-law, where the indices and the break
 momentum are identical for both accelerated protons and electrons.
$s_{\rm L}$ is the spectral index in momentum below the break.}
\tablenotetext{\rm c}{$p_{\rm br}$ is the break momentum.}
\tablenotetext{\rm d}{Spectral index in momentum above the break.}
\tablenotetext{\rm e}{Average hydrogen number density of ambient medium.}
\tablenotetext{\rm f}{The distance from the Earth is assumed to be 540~pc~\citep{Distance}. The total energy
 is calculated for particles $>$~100~MeV~$c^{-1}$.}
\tablenotetext{\rm g}{Seed photons for inverse Compton scattering of
 electrons include the CMB,
 two infrared~($T_{\rm IR} = 34, 4.7 \times 10^2$~K, $U_{\rm IR} = 0.34,
 6.3 \times 10^{-2}$~eV~cm$^{-3}$, respectively), and
 two optical components~($T_{\rm opt} = 3.6 \times 10^3, 9.9 \times
 10^3$~K, $U_{\rm opt} = 0.45, 0.16$~eV~cm$^{-3}$, respectively) in the
 vicinity of the Cygnus Loop.}
\end{center}
\end{table}

\end{document}